\documentclass[prl,twocolumn,showpacs,preprintnumbers,amsmath,amssymb]{revtex4}

\usepackage{graphicx}
\usepackage{dcolumn}
\usepackage{bm}

\begin{document}

\preprint{APS/123-QED}

\title{$^{16}{\rm O}$+$^{16}{\rm O}$ nature of the superdeformed band of
$^{32}{\rm S}$ \\
and the evolution of the molecular structure
}

\author{Masaaki Kimura}
\affiliation{RI-beam Science Laboratory, RIKEN (The Institute of
Physical and Chemical Reserch), Wako, Saitama 351-0198, Japan}
\author{Hisashi Horiuchi}
\affiliation{Department of Physics, Kyoto University, Kitashirakawa,
Kyoto 606-8502, Japan}


%

\begin{abstract}
The relation between the superdeformed band of $^{32}{\rm S}$ and
$^{16}{\rm O}$+$^{16}{\rm O}$ molecular bands is studied by the
deformed-base antisymmetrized molecular dynamics with the Gogny D1S
force. It is found that the obtained superdeformed band members of
$^{32}{\rm S}$ have considerable amount of the $^{16}{\rm O}$+$^{16}{\rm O}$ 
component. Above the superdeformed band, we have obtained two excited
 rotational bands which have more prominent character of the   
$^{16}{\rm O}$+$^{16}{\rm O}$ molecular band. These three rotational
bands are regarded as a series of $^{16}{\rm O}$+$^{16}{\rm O}$
molecular bands which were predicted by using the unique $^{16}{\rm O}$
-$^{16}{\rm O}$ optical potentil. As the excitation energy and principal
quantum number of the relative motion increase, the
$^{16}{\rm O}$+$^{16}{\rm O}$ cluster structure becomes more prominent
but at the same time, the band members are fragmented into several
states.    
\end{abstract}

\pacs{Valid PACS appear here}
\maketitle
The properties of the $^{16}{\rm O}$+$^{16}{\rm O}$ molecular bands have
been studied by many authors with the microscopic cluster models for
many years \cite{RGM}. Despite of these studies, the microscopic models
have not  been able to give a conclusive answer. One of the reasons is
the fact that the number of the molecuar band, the excitation energy of
the band head, and the moment of the inertia strongly depend on the
effective nuclear force. Recently, a rather conclusive answer was given
by the studies with the macroscopic model \cite{KONDO, OHKUBO}. In those
studies, the authors 
used the unique potical potential for the $^{16}{\rm O}$-$^{16}{\rm O}$
system which was determind in 1990's \cite{NICOLI} after the first
discovery of the nuclear rainbow in 1989 \cite{ORTZEN}. These studies
gave the following answers for the lowest three rotatinal bands whose
principal quantum numbers N=2n+L of the relative motion between clusters
are N=24, 26 and 28, respectively: The lowest Pauli-allowed rotational
band (N=24) starts from the $0^+$ state located at about 9 MeV in the
excitation energy (about 8 MeV below the $^{16}{\rm O}$+$^{16}{\rm O}$
threthold), and the energy gap between the N=24 and N=26 bands and that
between N=26 and N=28 bands are both approximately 10 MeV. In
Ref.\cite{OHKUBO}, it is proposed that the observed $^{16}{\rm
O}$+$^{16}{\rm O}$ molecular states correspond to the third band whose
principal quantum number is N=28.  

Besides the cluster models, in these days, the superdeformed structure
of $^{32}{\rm S}$ has been studied by many authors with the
mean-field theories \cite{YAMAGAMI, DOBACZEWSKI, RODRIGUEZ}. It is
largely because the superdeformed structure 
of $^{32}{\rm S}$ is regarded as a key to understand the relation
between the superdeformed state and the molecular structure. 
Indeed, by the HF(B) calculations, it is shown that the
superdeformed minimum of the energy surface is well established in
each angular momentum, and at the superdeformed local minimum, the wave
function shows the two-center-like character. It is also notable that
many of the mean-filed calculations predict that the superdeformed band
starts from the $0^+$ located at around 10 MeV which agrees with the
band head energy of the N=24 band obtained from the unique optical
potential. Therefore, it is enough conceivable that the superdeformed
band obtained by the mean-field calculations and the lowest
Pauli-allowed $^{16}{\rm O}$+$^{16}{\rm O}$ molecular band (N=24) are
identical.  

In the present study, we aim at clarifying the relation between the
superdeformed state and the $^{16}{\rm O}$+$^{16}{\rm O}$ molecular
structure. The objectives of the present study are summarized as
follows. (1) To what extent the superdeformed state and the $^{16}{\rm
O}$+$^{16}{\rm O}$ molecular structure are related? : In the unique
optical potential analysis, the factors which distort the $^{16}{\rm
O}$+$^{16}{\rm O}$ cluster structure such as the effects of the
spin-orbit force and the formation of the deformed one-body field are
not treated directly. Instead, these factors are renormalized into the
optical potential through the extrapolation to the low-energy
region. When one treats these factors veraciously by the microscopic
models, the pure $^{16}{\rm O}$+$^{16}{\rm O}$ cluster structure will
be distorted and will have a deformed one-body field character. In
other words, the superdeformed states in the mean-field models and
the states of the lowest Pauli-allowed $^{16}{\rm O}$+$^{16}{\rm O}$
band of the unique optical potential will be the states which have both
characters of the deformed one-body filed structure and two cluster
structure. Therefore, it will be important to study to what extent each
state has the $^{16}{\rm O}$+$^{16}{\rm O}$ character 
and the deformed one-body field character. (2) Do the excited states
exist in which the excitation energy is spent to excite the relative
motion between the clusters? Do they correspond to the N=26
and 28 bands obtained from the unique optical potential? : When
we believe that the superdeformed states of $^{32}{\rm S}$ have the
considerable $^{16}{\rm O}$+$^{16}{\rm O}$ component, we can expect the
excitation mode in which the excitation energy is used to excite the
relative motion between the clusters. 

The deformed-base antisymmetrized molecular dynamics (deformed-base AMD)
\cite{DEF_AMD}
combined with the generator coordinate method (GCM) has been used with
the Gogny D1S force \cite{GOGNY}. For the sake of the selfcontainedness,
we briefly explain this framework. For more details, the reader is
refered to the references \cite{AMDS}. The intrinsic basis wave function
of the system $\Phi_{int}$ is expressed by a Slater determinant of
single-particle wave packets $\varphi_i$. Each single-particle wave
packet is composed of spatial part $\phi_i$, spin part $\chi_i$ and
isospin part $\tau_i$. The spatial part has the form of the deformed
Gaussian centered at ${\bf Z}_i$.   
\begin{eqnarray}
 \Phi_{int}&=& \frac{1}{\sqrt{A!}}\det\{\varphi_i({\bf r}_j)\},\\
 \varphi_i({\bf r}_j) &=& \phi_i({\bf r}_j)\chi_i\tau_i,\\
 \phi_i({\bf r}) &=& \exp\{-\nu_x(x-{\rm Z}_{ix})^2-\nu_y(y-{\rm Z}_{iy})^2
  \nonumber\\ 
 &&-\nu_z(z-{\rm Z}_{iz})^2\},\\
 \chi_i &=& \alpha_i\chi_{\uparrow}+\beta_i\chi_{\downarrow}, \quad \tau_i = proton\ or\ neutron.
\end{eqnarray}
Here, the centroids of the Gaussian ${\bf Z}_i$ and the spin direction
$\alpha_i$ and $\beta_i$ are complex parameters and are dependent on
each particle. The width parameters $\nu_x$, $\nu_y$ and $\nu_z$ are
common to all particles. These variational parameters ($Z_i$,
$\alpha_i$, $\beta_i$ and ($\nu_x$,$\nu_y$, $\nu_z$)) are determind by
the variational calculation. The variational calculaton is made after
the parity projection by using parity-projected wave funciton
$\Phi^{\pi}=\frac{1\pm P_x}{2}\Phi_{int}$ as the variational 
wave function. In this study, the variational calculation is made under
the constraint on the nuclear deformation parameter $\beta$. The
advantage of the deformed Gaussian basis as the single-particle wave
packet is that it is possible to describe both the deformed 
one-body-field structure and the cluster structure as well as their
mixed structure within the same framework. We can confirm this feature
when we consider the two limits of the nuclear structure described
by this wave function, the deformed-harmonic-oscillator limit and the
cluster limit. The deformed-harmonic-oscillator limit is reached when
the centroids of all single-particle wave packets (Re ${\bf Z}_i$) are
at the center of the nucleus and the single-particle wave packets are
deformed. On the contrary, the cluster limit is obtained when the
centroids of the single-particle wave packets are separeted into the
centers of the constituent clusters and the single-particle wave packets
are spherical. 

After the constrained variational calculation for $\Phi^{\pi}$, we
superpose the optimized wave functions for various deformation
parameters (GCM): 
\begin{eqnarray}
 \Phi^{J^\pi} = c P^J_{MK} \Phi^{\pi}(\beta_0) + 
  c^{\prime} P^J_{MK^{\prime}} \Phi^{\pi}(\beta_0^{\prime}) + ...,
\end{eqnarray}
where $\Phi^{\pi}(\beta_0)$ is the optimized wave function under the
constraint of the nuclear deformation parameter $\beta=\beta_0$ and
$P^{J}_{MK}$ is the angular momentum projector. The cofficients $c,
c^{\prime}, ...$ are determined by the diagonalization of the
Hamiltonian. When the intrinsic wave function $\Phi^{\pi}(\beta_0)$ has
a prominent cluster structure, the superposition of the wave functions
will improve mainly the description of the relative wave function between
clusters, while it will improve mainly the description of the mean-field and
other correlations when $\Phi^{\pi}(\beta_0)$ has a prominent mean-field
character.

Below we explain how we estimate the amount of the $^{16}{\rm
O}$+$^{16}{\rm O}$ component in the obtained wave function
$\Phi^{J\pi}$. We rewrite the $\Phi^{J\pi}$ by decomposing it into the
$^{16}{\rm O}$+$^{16}{\rm O}$ component and the residual part,   
\begin{eqnarray}
 \Phi^{J\pi} = \alpha {\mathcal A}\{\chi_J(r){\rm Y}_{J0}(\hat r)\phi({^{16}{\rm O}})\phi({^{16}{\rm O}})\} + \sqrt{1-\alpha^2} \Phi_r,
\end{eqnarray}
where ${\mathcal A}$ is the antisymmetrizer, ${\bf r}$ is the relative
coordinate between two $^{16}{\rm O}$ clusters and $\phi({^{16}{\rm O}})$
is the internal wave function of the $^{16}{\rm O}$ cluster. $\chi_J(r)$
is so normalized that ${\mathcal A}\{\chi_J(r){\rm Y}_{J0}(\hat
r)\phi({^{16}{\rm O}})\phi({^{16}{\rm O}})\}$ is normalized to
unity. $\Phi_r$ is orthogonal to the $^{16}{\rm O}$+$^{16}{\rm O}$
cluster space. It is possible to evaluate the amount of the $^{16}{\rm
O}$+$^{16}{\rm O}$ component $w^J=|\alpha|^2 = |\langle
\Phi^{J\pi}|P^J_L|\Phi^{J\pi}\rangle|^2$ with the projection operator 
$P^J_L =\sum_\alpha|\widetilde{\varphi}^J_\alpha\rangle
\langle\widetilde{\varphi}^J_\alpha|$ onto the $^{16}{\rm O}$+$^{16}{\rm
O}$ cluster model space, where $\widetilde{\varphi}_\alpha^J$ is the
orthonormalized set of the spin projected Brink wave functions for
$^{16}{\rm O}$+$^{16}{\rm O}$ system. In this study, 26 spin-projected
Brink wave functions with the inter-cluster distance ranging from
0.5 fm to 13.0 fm are orthonormalized to construct $\widetilde{\varphi}_\alpha^J$.  

\begin{figure}
\includegraphics[width=\hsize]{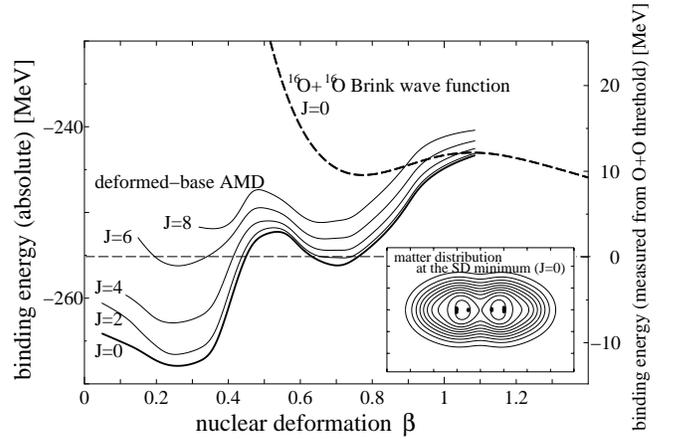}
\caption{The energy surface as a function of the nuclear deformation
 $\beta$ for $^{32}{\rm S}$. Dashed curve is for the $J=0$ state obtained
by the $^{16}{\rm O}$+$^{16}{\rm O}$ Brink wave function. Solid curves
are for J=0, 2, 4, 6 and 8 states obtained by the deformed-base AMD. The
matter density distribution of the deformed-base AMD wave function at
the superdeformed minimum (J=0) is also shown. Small black circles in
the density distribution represent the centroids of the single-particle
wave packets ${\rm Re}{\bf Z}_i$}
\label{FIG_SURFACE} 
\end{figure}

\begin{figure}
\includegraphics[width=\hsize]{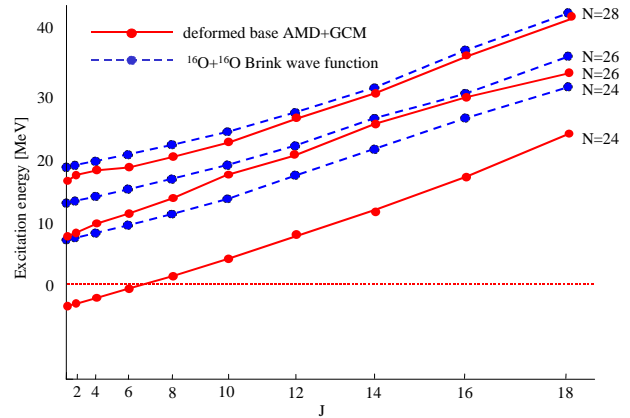}
\caption{The excitation energies of the N=24, 26 and 28 band members
obtained by the $^{16}{\rm O}$+$^{16}{\rm O}$ Brink wave function
(dashed lines) and the deformed-base AMD+GCM (solid lines). The N=26 and
 N=28 band members are fragmented into several states in the
 deformed-base AMD+GCM calculation and the averaged energies
 ${\rm E}_{\rm AV}$ are  shown for these bands.} 
\label{FIG_BAND_1} 
\end{figure}

Before discussing the results of the deformed-base AMD+GCM, we
show the result of the calculation in which we use the intrinsic
wave functions which have the pure $^{16}{\rm O}$+$^{16}{\rm O}$
configuration. The Brink wave function is used for the intrinsic
$^{16}{\rm O}$+$^{16}{\rm O}$ wave function. Therefore, this result
is the same as the traditional cluster model calculation of $^{16}{\rm
O}$+$^{16}{\rm O}$ RGM (rosonating group method) and $^{16}{\rm
O}$+$^{16}{\rm O}$ GCM. In Fig. \ref{FIG_SURFACE}, the
energy surface for the $J=0$ state as a function of the nuclear
deformation is shown (dashed curve). We note that the $^{16}{\rm
O}$+$^{16}{\rm O}$ configuration describes more than $4\hbar\omega$
excited states relative to the ground state of $^{32}{\rm S}$, therefore
the ground state and the normal deformed states are not included in this
energy surface. It has the energy minimum at $\beta=0.73$
(inter-cluster distance is 5.0 fm) which corresponds to two touching
$^{16}{\rm O}$. The minimum energy is about 8 MeV higher than the
$^{16}{\rm O}$+$^{16}{\rm O}$ threthold energy. After the GCM
calculation along this energy surface, we have obtained three rotational
bands which have the quantum numbers of the relative motion N=24, 26 and
28, respectively, and are shown by dashed lines in
FIG. \ref{FIG_BAND_1}. However, their energies are too high to coincide
with  the rotational bands obtained from the unique optical potential
and also with the superdeformed band obtained from the HF(B)
calculations. The energy gaps between these bands (4 MeV between N=24
and N=26, and 6 MeV between N=26 and N=28 in the case of $0^+$ states)
are much smaller than the results from the unique optical potential. We
think that these deviations come from the fact that the effects which
distort the cluster structure are neglected in the Brink wave
function. We will see below that in fact the effects of the distortion
are fairly large. 

\begin{table*}
 \caption{Excitation energy Ex [MeV] and the intra-band $E2$ transition
 probabilities B($E2$;$J\rightarrow J-2$) [${\rm e^2fm^4}$] of the ground band
 and the first excited band.}\label{TAB_LOWLYING}
\begin{ruledtabular}
\begin{tabular}{ccccccccccc}
 &\multicolumn{2}{c}{ground (Theor)}
 &\multicolumn{2}{c}{ground (Exp)}
 &\multicolumn{2}{c}{band I (Theor)}
 &\multicolumn{2}{c}{band I (Exp)}\\
$J$ & Ex & B(E2)& Ex & B(E2)& Ex & B(E2)& Ex & B(E2)\\
\hline
 0  &  - & -   &  -   &  -   &  3.9   &  -    & 3.778   &  -    \\
 2  & 2.3& 66  & 2.23 & 60$\pm$6 &  4.8   &  31   &  4.282  &  -    \\
 4  & 5.75 & 109  &4.459  & 72$\pm$12  &  10.1   &  88  & 6.852   & 35.4$^{+18.6}_{-8.4}$     \\
 6  & 10.2 & 130 & 8.346  & $>22.2$  &  12.9   &  98   & 9.783   &  -    \\
\end{tabular}
\end{ruledtabular}
\end{table*}

 In Fig. \ref{FIG_SURFACE}, the energy surfaces for $J=0$ to $J=8$ states
obtained by the deformed-base AMD+GCM calculation are also shown. Since
the AMD wave function does not assume any cluster configuration, the
normal deformed states also appear in these energy surfaces. In the
normal deformed states, the prolately deformed states and the triaxially
deformed states ($\gamma=6^{\circ}\sim 30^{\circ}$) are energetically 
degenerate. After the GCM 
calculation, prolate wave functions mainly contribute to the ground band 
while the triaxially deformed wave functions to the first excited
band. Their excitation energies and the intra-band $E2$ trainsition
probabilities (Table \ref{TAB_LOWLYING}) show reasonable 
agreement with experiments and are consistent with the results of the
HFB calculation with the Gogny D1S force \cite{RODRIGUEZ}, though the
total binding energy of the ground state underestimates the experimental
data by about  2.0 MeV. 

We also find that the behavior of the energy surface around the excited
local minimum is similar to that of the HF(B) calculations. In each
angular momentum state, the superdeformed mimimum is well developed and
the $0^+$ excitation energy relative to the normal deformed ground state
is around 10 MeV. The energy difference between the deformed-base AMD
and the $^{16}{\rm O}$+$^{16}{\rm O}$ calculation at the superdeformed
minimum is about 10 MeV, which implies a fairly large effect of the
distortion of the cluster structure. Indeed, the deformed-base AMD wave
function deviate from the pure cluster limit. At the superdeformed
limit, the single-particle wave packets are prolately deformed
($\nu_x=\nu_y=0.160$ ${\rm fm}^{-2}$ and $\nu_z=0.115$ ${\rm fm}^{-2}$),
and the distance between the centroids of the single-particle wave
packets are rather small (3.1 fm), though they are separated into two
parts (centers of two $^{16}{\rm O}$) and show two-center nature. The
energy gain by the deformed-base AMD function compared to the pure
$^{16}{\rm O}$+$^{16}{\rm O}$ wave function mainly comes from the
two-body spin-orbit force and the density dependent force. In the case
of the deformed-base AMD wave function, the expectation value of the
two-body spin-orbit force is about -4.5 MeV which must be zero in the
pure $^{16}{\rm O}$+$^{16}{\rm O}$ wave function. Though its value is
not so large, it lowers the excitation energy of the superdeformed state
from the $^{16}{\rm O}$+$^{16}{\rm O}$ limit. Compared to the $^{16}{\rm
O}$+$^{16}{\rm O}$ wave function, the expectation value of the repulsive
density dependent force is about 6 MeV smaller in the deformed-base AMD
wave function and it also indicates the non-small deviation from the 
$^{16}{\rm O}$+$^{16}{\rm O}$ structure. Though the kinetic energy does
not much contribute to lower the energy of the superdeformed band, its
nature is different from the case of the pure $^{16}{\rm O}$+$^{16}{\rm
O}$ structure. At the superdeformed minimum, the single-particle wave
packets are prolately deformed and since the kinetic energy is almost
linear to width parameter $\nu$, the kinetic nergy to the z direction is
eased. However, we have found that the wave function of the
deformed-base AMD (which is parity and the angular momentum projected)
at the superdeformed minimum still has a considerable amount of the
$^{16}{\rm O}$+$^{16}{\rm O}$ component, $w^{J=0}=0.57$. 

\begin{figure}
\includegraphics[width=\hsize]{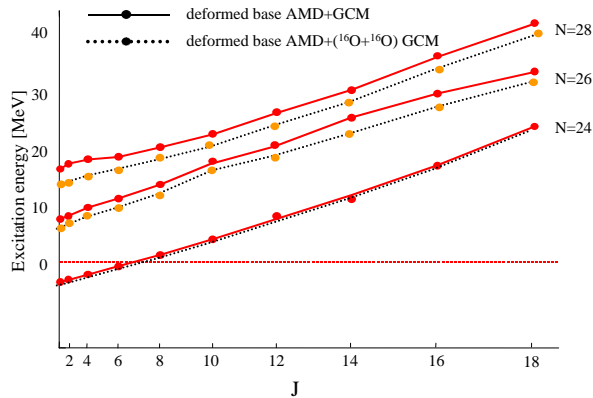}
\caption{The excitation energies of the N=24, 26 and 28 band members
obtained by the deformed-base AMD+GCM (solid lines) and the
 deformed-base AMD+($^{16}{\rm O}$+$^{16}{\rm O}$)+GCM (dotted lines).
 The N=26 and N=28 band members are fragmented into several states in
 both calculations and the averaged energies ${\rm E}_{\rm AV}$ are
 shown for these bands. The deforemd-base AMD+GCM results in this figure
 are the same as those of FIG. \ref{FIG_BAND_1}.}
\label{FIG_BAND_2} 
\end{figure}
After the GCM calculation by superposing the deformed-base AMD wave
functions along the energy surface, we have obtained three rotational
bands. The lowest superdeformed band, the second lowest band and the
third band have large $^{16}{\rm O}$+$^{16}{\rm O}$ components with
N=24, 26 and 28, respectively. Thus these three bands correspond to the
N=24, 26 and 28 bands of the unique optical potential. We refer to these
bands simply as N=24, 26 and 28 bands (Fig. \ref{FIG_BAND_1}). The band
members of the N=26  and 28 bands are fragmented into several
states. Namely, there are several states which have the $^{16}{\rm
O}$+$^{16}{\rm O}$ component of the same principal quantum number N.  
For example, the $0^+$ state of the N=26 band is fragmented into three
states in which $w^{J=0}$ are 0.34, 0.13, and 0.25, respectively. In the
following, we only show the averaged 
energies of these fragmented states to investigate the gross feature of
the band structure and the fragmentation is discussed later. The
averaged energy is calculated by multiplying the overlap $w_i$ as the
weight; $E^J_{\rm AV}=(w^J_1E^J_1+w^J_2E^J_2+...)/(w^J_1+w^J_2+...)$.    
The excitation energies of the N=24 band members are lowered by about
3 MeV compared to those of the band formed by the states at the
superdeformed energy minimums. This may be due to the improvement of the 
description of the mean-field and the collective motion by the GCM
calculation, since in the N=24 band, the amount of the $^{16}{\rm
O}$+$^{16}{\rm O}$ component is decreased by the GCM calculation;
$w^{J=0}=0.57$ for the deformed-base AMD wave function and
$w^{J=0}=0.42$ for the deformed-base AMD+GCM wave function.
It is important that the energy gain
of the deformed-base AMD+GCM band compared to the pure $^{16}{\rm
O}$+$^{16}{\rm O}$ band becomes smaller as the excitation energy of the
relative motion between clusters becomes larger. For example, compared
to the pure $^{16}{\rm O}$+$^{16}{\rm O}$ bands, the $0^+$ states of
N=24, 26 and 28 bands are lowered by about 12, 6 and 2 MeV,
respectively. This decrease of the energy gain means the enhancement of
the $^{16}{\rm O}$+$^{16}{\rm O}$ molecular structure in the higher
excited bands. Indeed, the sum of the $w^J$ of the fragmented band  
members drastically increases in N=26 and 28 bands compared to that
of the N=24 band; they amount to 0.71 and 0.73 for the case of the $0^+$
states of the N=26 and 28 bands. 

In the framework of the deformed-base AMD+GCM, the basis states for the
GCM calculation are obtained from the variational calculation. Since the 
variational calculation optimizes mainly the lowest N=24 band
members in which the $^{16}{\rm O}$+$^{16}{\rm O}$ molecular structure
is distorted, these basis states may be inappropriate to describe the
N=26 and N=28 bands in which the $^{16}{\rm O}$+$^{16}{\rm O}$ molecular
structure is drastically enhanced. In other words, the present
calculation may underestimate the enhancement of the $^{16}{\rm
O}$+$^{16}{\rm O}$ molecular structure in N=26 and N=28
bands. Therefore, we have included the $^{16}{\rm O}$+$^{16}{\rm O}$ Brink
wave functions in the basis states of the GCM calculation in addition to
the deformed-base AMD wave functions obtained from the variational
calculation. The obtained results of the enlarged GCM calculation are
shown in Fig. \ref{FIG_BAND_2}. The enhancement of the $^{16}{\rm
O}$+$^{16}{\rm O}$ molecular structure has become more prominent in N=26
and N=28 bands. It is reasonable that the excitation energies and the
overlaps $w^J$ of the N=24 band members do not change much, since in
this band, the $^{16}{\rm O}$+$^{16}{\rm O}$ molecular structure is
distorted and the inclusion of the pure $^{16}{\rm O}$+$^{16}{\rm O}$
configuration is less important. On the contrary, in the N=26 and 28
bands, the excitation energies are lowered by about a few MeV and the
amounts of the $^{16}{\rm O}$+$^{16}{\rm O}$ component have increased
drastically. The sums of the fragmented $w^J$ for N=26 and 28 bands are
0.90 and 0.98, respectively. It is interesting that the highest N=28 band
members have almost pure $^{16}{\rm O}$+$^{16}{\rm O}$ molecular
structure when we sum up the fragmented states, and it looks plausible
that this band members are assigned to correspond to the observed
molecular resonances of $^{16}{\rm O}$+$^{16}{\rm O}$ as is proposed in
Ref.\cite{OHKUBO}.  

\begin{table}
 \caption{The excitation energies $E_x$ and the amounts $w^J$ of the 
 $^{16}{\rm O}$+$^{16}{\rm O}$ components of the fragmented $0^+$
 states of N=26 and 28 bands.}
\label{TAB_FRAGMENT} 
\begin{ruledtabular}
\begin{tabular}{c|cccc}
 &\multicolumn{2}{c}{N=26}&\multicolumn{2}{c}{N=28}\\
 &$E_x$& $w^J$ &$E_x$& $w^J$\\
\hline
fragment I & 23.8 & 0.54  & 31.2   & 0.32  \\
fragment II  & 24.0 & 0.13  & 34.0  & 0.45  \\
fragment III  & 25.3  & 0.23  & 38.7  & 0.20     \\
${\rm E}_{\rm AV}$ and $\sum w^J$ &  24.2  & 0.90  & 33.7    & 0.98\\
\end{tabular}
\end{ruledtabular}
\end{table}

Finally, we discuss the fragmentation of the N=26 and 28 band
members. The judgement whether a fragment belongs to the N=26 or 28 band
is determined by evaluating the amount of the overlaps with the N=26 and
28 band members obtained from the GCM calculation with the pure
$^{16}{\rm O}$+$^{16}{\rm O}$ wave function.
As an example, the energies and the amounts of the $^{16}{\rm
O}$+$^{16}{\rm O}$ component for $0^+$ fragments are listed in Table
\ref{TAB_FRAGMENT}. The $0^+$ states of the N=26 and 28 bands are
fragmented into three states. In the present calculation, the number of
the fragments do not strongly depend on the angular momentum and the
principal quantum number N. At most, the $12^+$ state of N=28 band is
fragmented into five states. These fragmentations are mainly caused by
the coupling with the states with medium deformation  which appear as a
small peak between normal deformed states and the superdeformed states
and also by the coupling with the $^{16}{\rm O}$+$^{16}{\rm O}^*$ states
which are included in the largely deformed ($\beta>0.9$) wave functions
where $^{16}{\rm O}^{*}$ stands for distorted $^{16}{\rm O}^{*}$
cluster. Details of these couplings are important to compare the present 
results and  the experiments and will be investigated in the future.

To summarize, we have shown that the superdeformed band obtained from
the HF(B) calculations and the Pauli-allowed lowest N=24 band of the
$^{16}{\rm O}$+$^{16}{\rm O}$ molecular bands are essentially
identical. In this band, $^{16}{\rm O}$+$^{16}{\rm O}$ molecular
structure is distorted by the effects of the deformed mean-field
formation and the spin-orbit force. This distortion is not small and
lowers the excitation energy largely, but this band members still have
the considerable component of the $^{16}{\rm O}$+$^{16}{\rm O}$
molecular structure. We have obtained two excited bands which are
generated by the excitation of the relative motion between two
$^{16}{\rm O}$ clusters contained in the $^{16}{\rm O}$+$^{16}{\rm O}$
component of the superdeformed band. In the excited N=26 and 28 bands,
the distortion is less important and the band members have the prominent
molecular structure. The members of these bands are fragmented into
several states and the assigment of the N=28 band members to the oberved
$^{16}{\rm O}$+$^{16}{\rm O}$ molecular resonances looks plausible.

\begin{acknowledgments}
The authors would like to thank Dr. Y. Kanada-En'yo for useful
discussions. Most of the computational calculations were carried out by 
SX-5 at Reserch Center for Nuclear Physics, Osaka University (RCNP). 
This work was partially performed in the gResearch Project for Study of
Unstable Nuclei from Nuclear Cluster Aspectsh sponsored by Institute of
 Physical and Chemical Research (RIKEN). 
\end{acknowledgments}

\end{document}